\documentclass[aps,apl,preprint,superscriptaddress]{revtex4-1}
\usepackage{graphicx}
\usepackage{amsmath}
\usepackage{mathtools}
\usepackage{amssymb}
    \DeclareMathOperator{\sech}{sech}

\begin{document}

\title{Anomalous direction for skyrmion bubble motion}

\author{Fanny C. Ummelen$^*$}
\affiliation{Department of Applied Physics, Center for NanoMaterials and COBRA Research Institute, Eindhoven University of Technology, P.O.~Box~513, 5600 MB Eindhoven, The Netherlands}
\author{Tijs A. Wijkamp}
\affiliation{Department of Applied Physics, Center for NanoMaterials and COBRA Research Institute, Eindhoven University of Technology, P.O.~Box~513, 5600 MB Eindhoven, The Netherlands}
\author{Tom Lichtenberg}
\affiliation{Department of Applied Physics, Center for NanoMaterials and COBRA Research Institute, Eindhoven University of Technology, P.O.~Box~513, 5600 MB Eindhoven, The Netherlands}
\author{Rembert A. Duine}
\affiliation{Institute for Theoretical Physics, Utrecht University, Leuvenlaan 4, 3584 CE Utrecht,
The Netherlands}
\affiliation{Department of Applied Physics, Center for NanoMaterials and COBRA Research Institute, Eindhoven University of Technology, P.O.~Box~513, 5600 MB Eindhoven, The Netherlands}

\author{Bert Koopmans}
\affiliation{Department of Applied Physics, Center for NanoMaterials and COBRA Research Institute, Eindhoven University of Technology, P.O.~Box~513, 5600 MB Eindhoven, The Netherlands}
\author{Henk J.M. Swagten}
\affiliation{Department of Applied Physics, Center for NanoMaterials and COBRA Research Institute, Eindhoven University of Technology, P.O.~Box~513, 5600 MB Eindhoven, The Netherlands}
\author{Reinoud Lavrijsen}
\affiliation{Department of Applied Physics, Center for NanoMaterials and COBRA Research Institute, Eindhoven University of Technology, P.O.~Box~513, 5600 MB Eindhoven, The Netherlands}

\date{\today}

\begin{abstract}
Magnetic skyrmions are localized topological excitations that behave as particles and can be mobile, with great potential for novel data storage devices. In this work, the current-induced dynamics of large skyrmion bubbles is studied. When skyrmion motion in the direction opposite to the electron flow is observed, this is usually interpreted as a perpendicular spin current generated by the spin Hall effect exerting a torque on the chiral N\'{e}el skyrmion. By designing samples in which the direction of the net generated spin current can be carefully controlled, we surprisingly show that skyrmion motion is always against the electron flow, irrespective of the net vertical spin-current direction. We find that a negative bulk spin-transfer torque is the most plausible explanation for the observed results, which is qualitatively justified by a simple model that captures the essential behaviour. These findings demonstrate that claims about the skyrmion chirality based on their current-induced motion should be taken with great caution.
\end{abstract}

\maketitle

Magnetic skyrmions are swirling spin textures that are currently the focus of many research efforts \cite{review_skyrmions}. In such a whirl the magnetization direction in the centre is opposite to the magnetization direction at the edge, while in-between the magnetization direction rotates gradually with a certain chirality. The interaction that favours this chiral rotation is the Dzyaloshinskii-Moriya interaction (DMI) \cite{dmi_original,dmi_original2}, which plays an important role in the creation and stabilization of magnetic skyrmions. This interaction is only present in systems with strong spin-orbit coupling (SOC) and structural inversion asymmetry (SIA), which occurs in certain crystals lacking inversion symmetry or in ultrathin multilayers near interfaces. In both type of systems with SIA, skyrmions have been observed experimentally \cite{skyrmion_observed1,skyrmion_observed2,skyrmion_observed3,klaui,skyrmionir}, and in this work we will focus on the latter. We observe a unique motion of skyrmion bubbles, that cannot be described by conventional effects.

\begin{figure}[t!]
\centering
\includegraphics[width=0.7\textwidth]{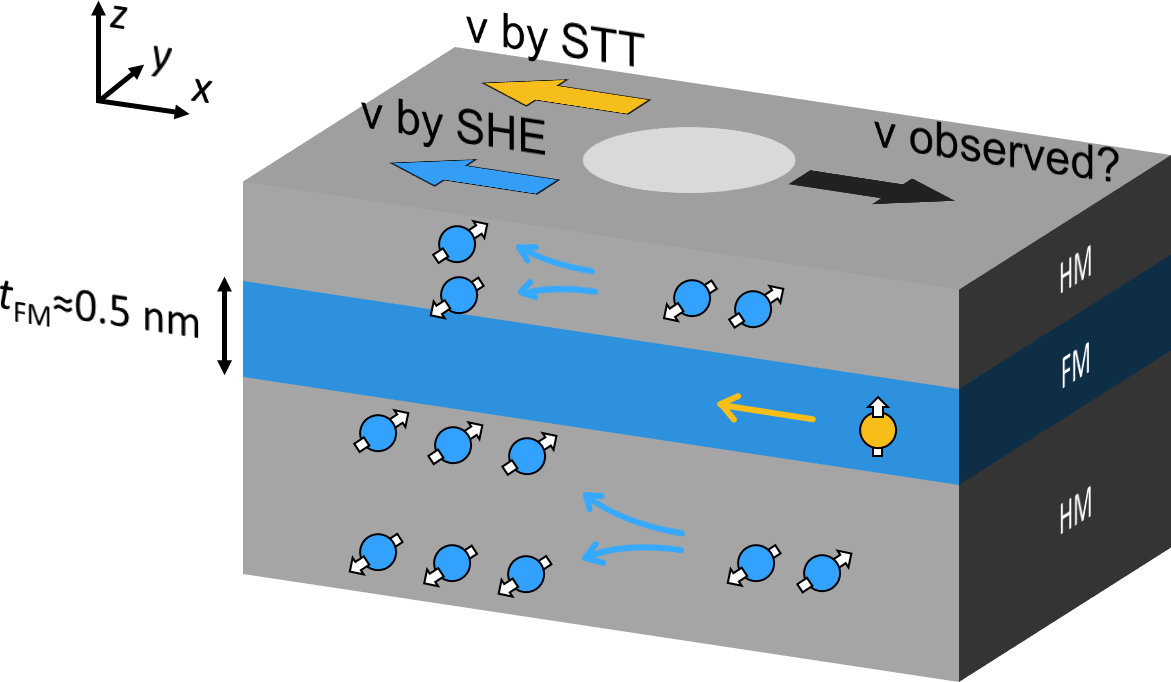}
\caption{Schematic view of the studied geometry: a thin FM layer sandwiched between HM layers in which a skyrmion bubble is stabilized. Also the spin currents that can cause a torque on the skyrmion are depicted: a vertical spin current caused by the SHE and an in-plane spin current from conduction electrons that become spin polarized when flowing through a magnetic material. }
\label{C6:fig0}
\end{figure}

In order to become a viable candidate for future data carriers, the movement of skyrmions should be controlled, preferably by electrical current. Fig.\ \ref{C6:fig0} schematically shows a typical stack like we investigate in this work \textemdash an ultrathin ferromagnetic (FM) layer in between two heavy metal (HM) layers \textemdash and the current-induced effects that can propagate a skyrmion. 
Electrons moving in the plane of the sample through the magnetic layer become spin polarized, and exert a bulk spin-transfer torque (STT) on the skyrmion (indicated in yellow). This torque results from angular momentum conservation \cite{SLONCZEWSKI1996L1,Thiaville2005}, and in principle always leads to movement in the direction of the electron flow. The spin Hall effect (SHE) occurs in a heavy metal where a spin accumulation is generated perpendicular to the charge current (indicated in blue). If there is a magnetic material adjacent to this heavy metal the injected spins can exert a torque on the magnetization texture \cite{Hirsch1999,pascal}. This can lead to fast and efficient motion either along or against the electron flow, depending on the sign of the spin current and the orientation of the magnetization. In practice, heavy metal layers are generally present in material stacks that contain skyrmions, as they are also required to induce a sizable DMI. Also, bulk STTs are widely believed to be relatively small in ultrathin layers \cite{cormier}. Therefore it is not surprising that the SHE is considered to be the dominant driving mechanism in experimental reports on current-induced skyrmion motion, especially when motion against the electron flow is observed.

For the creation of samples that contain skyrmions, the approach of Schott \textit{et al.}, in which a wedge of magnetic material is investigated, is used \cite{schott}. In such samples, the skyrmions typically appear at small ($\sim 0.5$ nm) magnetic layer thicknesses, at the onset ferromagnetism. In this work, we systematically vary the vertical spin current injected by the SHE. Its size and sign are controlled by the thicknesses of the HM layers adjacent to the FM layer, by which we are able to control the expected direction of motion. We observe an unexpected, unique skyrmion motion that cannot be explained by the SHE or the traditional bulk STT, see the black arrow in Fig.\ \ref{C6:fig0}. In similar atomically thin magnetic layers, depinning measurements on magnetic domain walls (DWs) could only be explained by a \textit{negative} spin transfer torque \cite{negativeSTT}. Using a qualitative model, we show that this novel ingredient is a plausible explanation for the observed anomalous skyrmion motion.

\section{Sample design}

The perpendicular spin current density, $J_\textrm{S}$, induced by the SHE in a heavy metal layer is determined by $J_\textrm{S}=\theta_{\textrm{SH}} J$, where $\theta_{\textrm{SH}}$ is the spin Hall angle and $J$ the charge current density in the heavy metal. If the thickness, $t$, of the heavy metal layer is comparable to spin diffusion length, $\lambda_\textrm{sf}$, $J_\textrm{S}$ is reduced by a factor $1-\sech{\frac{t}{\lambda_{\textrm{sf}}}}$ \cite{sech}. This implies that in a Pt/Co/Pt sample the net injected spin current can be controlled by varying the thicknesses of both Pt layers \cite{pascal}, provided that $\theta_{\textrm{SH}}$ is the same for the top and bottom Pt layer. First, we verified this by showing that the net spin current is opposite for a Pt (4 nm)/ Co (0.6 nm) / Pt (2 nm) and a Pt (2 nm) / Co (0.6 nm) / Pt (4 nm) stack, see supplementary material. For the main experiment, a sample will be grown in which one of the Pt layers gradually varies in thickness. We choose to vary the top Pt layer thickness, so that the bottom Pt layer remains constant, which limits changes in properties determined by the subsequent interfaces, like the magnetic anisotropy or DMI. The thickness variation is chosen such that both the situation in which the dominant contribution from the SHE comes from the bottom Pt layer, and the situation in which the dominant contribution comes from the top Pt layer can be studied. For purely SHE driven motion, the direction of motion is expected to be opposite in these two regimes.

\begin{figure}[t!]
\centering
\includegraphics[width=0.6\textwidth]{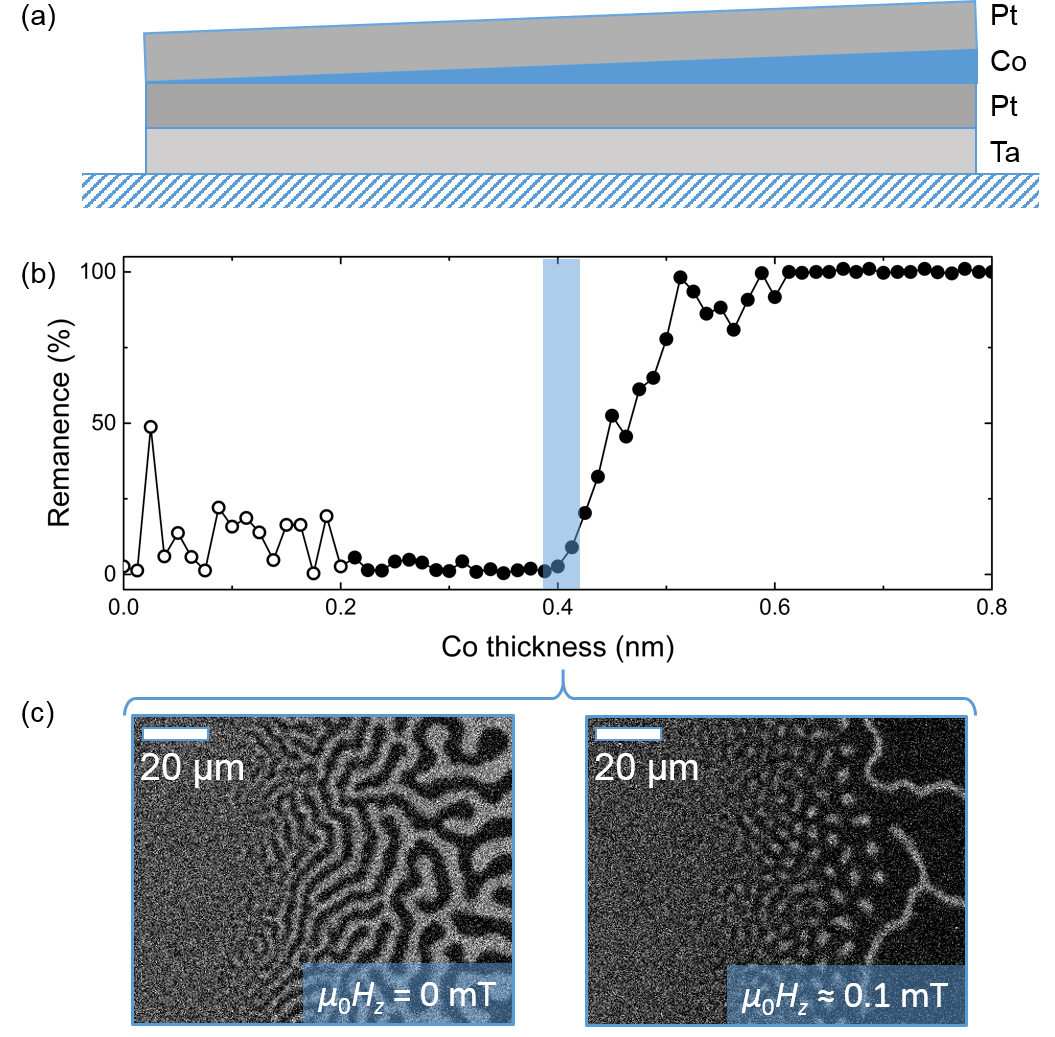}
\caption{(a) Schematic representation of the studied sample. (b) Remanence of the sample as a function of the Co thickness. A transition where the PMA disappears is observed. The non-zero points below 0.2 nm (indicated by open circles) have no physical meaning. They occur because at these thicknesses only noise is measured, leading to problems in the algorithm determining the remanence. (c) Magnetic domain structures observed by Kerr microscopy at the transition, both with and without the application of a small perpendicular magnetic field.}
\label{C6:fig1}
\end{figure}

Besides controlling the direction of the net spin current, it is necessary for the experiment to stabilize skyrmion bubbles in the sample. Following the approach by Schott \textit{et al.}, the thickness of a magnetic layer is gradually varied. This results in a large variation of material parameters (especially the saturation magnetisation, the magnetic anisotropy and DMI are expected to vary with thickness) in one sample, allowing access to the narrow parameter window in which skyrmion bubbles can be stabilized \cite{schott}. Before addressing the aforementioned samples in which also the thickness of the adjacent Pt layers is varied, we first study samples in which only the magnetic layer varies in thickness. This is to check whether we can actually create skyrmion bubbles in (almost) symmetric material stacks, in which the direction of the net spin current can be controlled. The used material stack is Ta (5 nm) / Pt (4 nm) / Co(0-0.8 nm) / Pt (4 nm), schematically shown in Fig.\ \ref{C6:fig1}(a).

The remanence as a function of the Co layer thickness is obtained using polar MOKE measurements and is shown in Fig.\ \ref{C6:fig1}(b). In the graph two regimes can be identified: for Co thicknesses above 0.5 nm the sample has an out-of-plane easy axis and the remanence is 100 percent. For thicknesses below 0.4 nm the remanence is zero, because the ferromagnetic-paramagnetic transition is approached and the ferromagnetic state disappears. When the remanence turns to zero at 0.4 nm Co thickness, the magnetization and out-of-plane easy axis do not disappear immediately (though the values of $M_\textrm{S}$ and $K_\textrm{eff}$ are significantly reduced \cite{bubbles_Os,schott}): a small regime exists in which at zero applied magnetic field multiple small out-of-plane magnetized domains appear, and the up and down domain are equal in size. This results in a measured remanence of zero, since in this measurements we average over a large area on the sample containing many up and down domains. Using Kerr microscopy a labyrinth structure is observed at remanence in this regime, but when a small magnetic field ($\sim 0.1$ mT) is applied circular, micron-sized domains spontaneously form, see  Fig.\ \ref{C6:fig1}(c). These are the skyrmion bubbles that will be studied in the remainder of this paper, and the thickness at which they occur will be referred to as the skyrmion transition.

In the supplementary material the characterization of the skyrmion bubbles is shown (here also the results for Pt/Co/Ta and Pt/Co/Ir samples are shown), and the general conclusions are the following: for all investigated stacks the diameter of the bubbles is of the order of $1$ $\mu$m, the field at which they are stabilized is of the order of $0.1$ mT and the Co thickness range in which they are stable is only 0.01 nm wide. This last value seems extremely low, as it is an order of magnitude smaller than the size of a single atom. First, one should note that in these sputter deposited systems all mentioned thicknesses are $nominal$ thicknesses, which do not have to be a multiple of the atomic layer thickness. Secondly, around the ferromagnetic-paramagnetic transition, the material parameters vary extremely rapidly with Co thickness (see for instance the supplementary material, where the DW energy as a function of thickness is determined). Because for the stabilization of skyrmions a delicate balance between material parameters is required, it is not surprising that they are only stable in a very narrow thickness range. So although the stability range is very narrow, the skyrmion transition was reproducibly found in all material stacks investigated, indicating that this method is a reliable way to create skyrmion bubbles, even in sample stack in which the magnetic layer is interfaced with the same material on both sides. This is a non-trivial result, as the DMI, which is believed to be essential in the stabilization of skyrmions, is typically small in such nominal symmetric samples (though not necessarily zero, because differences in growth can also lead to asymmetry \cite{bubble_reinoud}).

Considering the above, a sample to investigate the role of the SHE on skyrmion bubble motion is designed. This is a Pt (4 nm) / Co (0 - 0.6 nm) / Pt (0 - 10 nm) sample in a double wedge geometry, schematically shown in Fig.\ \ref{C6:fig3}(a). In one direction the Co thickness is varied, which is necessary to hit the narrow thickness range in which the skyrmion bubbles are stable. In the other direction, the thickness of the top Pt layer is varied, which ensures a gradually varying SHE induced spin current along the sample.

\section{Results}

\begin{figure}[t!]
\centering
\includegraphics[width=\textwidth]{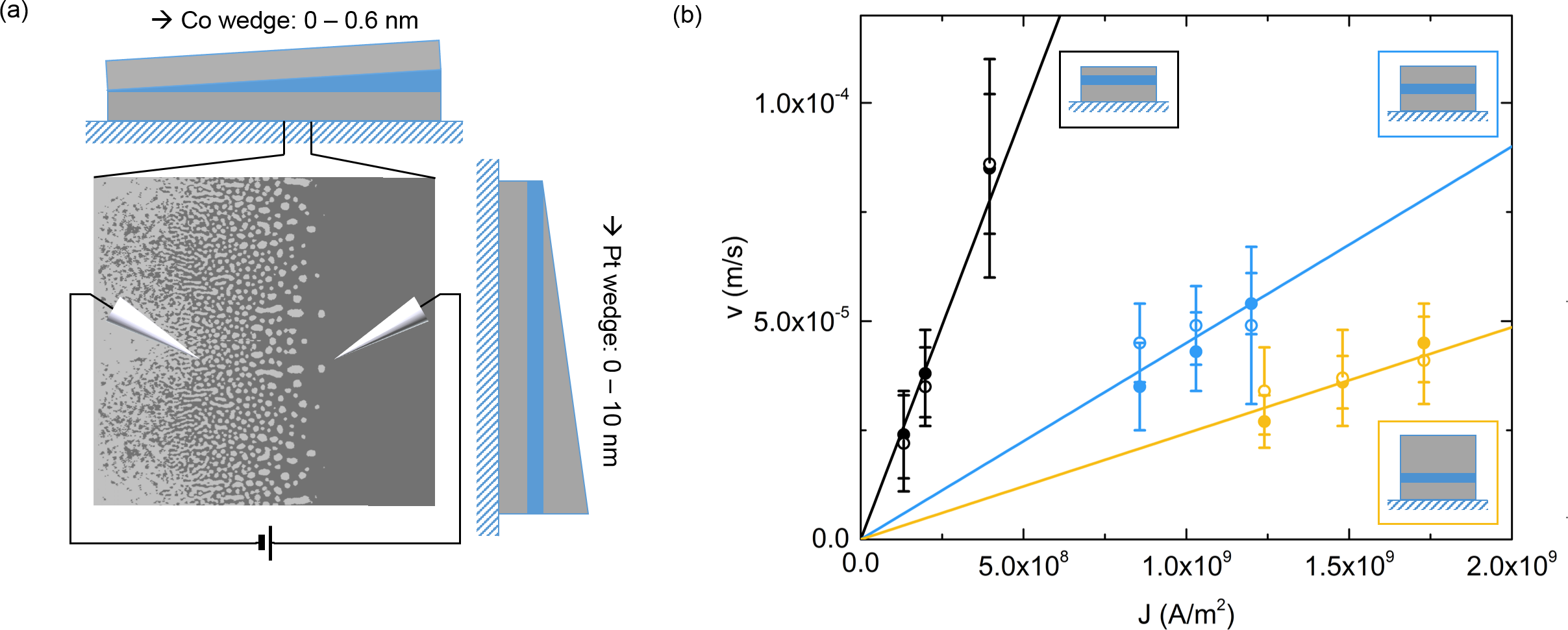}
\caption{(a) Schematic representation of the double wedge sample and the application of current. (b) Measured skyrmion bubble velocity at various thicknesses of the top Pt layer: 0.5 nm (black points), 4.5 nm (blue points) and 8.5 nm (yellow points). At each thickness three current densities are used, and skyrmion bubbles of both polarities (indicated by the open and closed points) are investigated. The lines are linear fits fixed through the origin. The large error bars are a consequence of the rapidly varying material parameters along the Co wedge, see supplementary material for further explanation.}
\label{C6:fig3}
\end{figure}

To perform the experiment, probes are landed around the skyrmion bubbles at each Pt thickness we want to investigate (steps of 2/3 nm thickness variation are taken) for current injection. The probes are typically placed 200 $\mu$m apart and the calculation of the current density and discussion on the homogeneity can be found in the supplementary information. At each Pt thickness 6 measurements are performed: three different current densities are investigated for both polarities \textemdash so both the case that the magnetization in the centre of the skyrmion bubble points in the +$z$ direction while the magnetization of the surroundings points in the -$z$ direction and vice versa \textemdash of the skyrmion bubbles. Fig.\ \ref{C6:fig3}(b) shows three typical measurements of the average velocity, $v$, as a function of the applied current density on this sample. Of the three data sets, the black points represent the situation in which the top Pt layer is much thinner than the bottom layer (1.5 nm), blue shows the situation where the Pt thicknesses are comparable (4.5 nm) and yellow shows the situation where the top layer is much thicker than the bottom layer(7.5 nm). Several features stand out: firstly, all measured velocities are positive, corresponding to movement along the charge current direction. This indicates that, surprisingly, no cancellation point or reversal of movement direction is observed. Secondly, with increasing top Pt thickness, larger current densities are required to reach the same velocities. Last, there seems to be a linear increase in velocity with the applied current.

At this point, we conclude that it is not possible to explain our observation by the SHE alone: in that case the direction of motion would be opposite for the situation with a bottom Pt layer that is thicker than the top Pt layer and vice versa. This might be explained by a difference in $\theta_{\textrm{SH}}$ or interface transparency for the top and bottom Pt layers, but additional experiments in the supplementary information and earlier work by Haazen \textit{et al.} \cite{inversion_asymmetry_growth,pascal}, render this unlikely. The conventional bulk spin-transfer torque also does not provide an explanation, as it predicts motion in the direction of the electron flow only. Je \textit{et al.} \cite{negativeSTT} already encountered a similar observation in an experiment on DWs and how efficient they are depinned by current. They explained their results by the presence of a \textit{negative} bulk spin transfer torque. We will now investigate if this is also a plausible explanation for our observations. To do this, we will use a model which includes both this negative bulk spin transfer torque and the SHE to qualitatively predict how the skyrmion velocity will change along the double wedge sample, and compare this to the experimental results.

 \begin{figure}[t!]
\centering
\includegraphics[width=\textwidth]{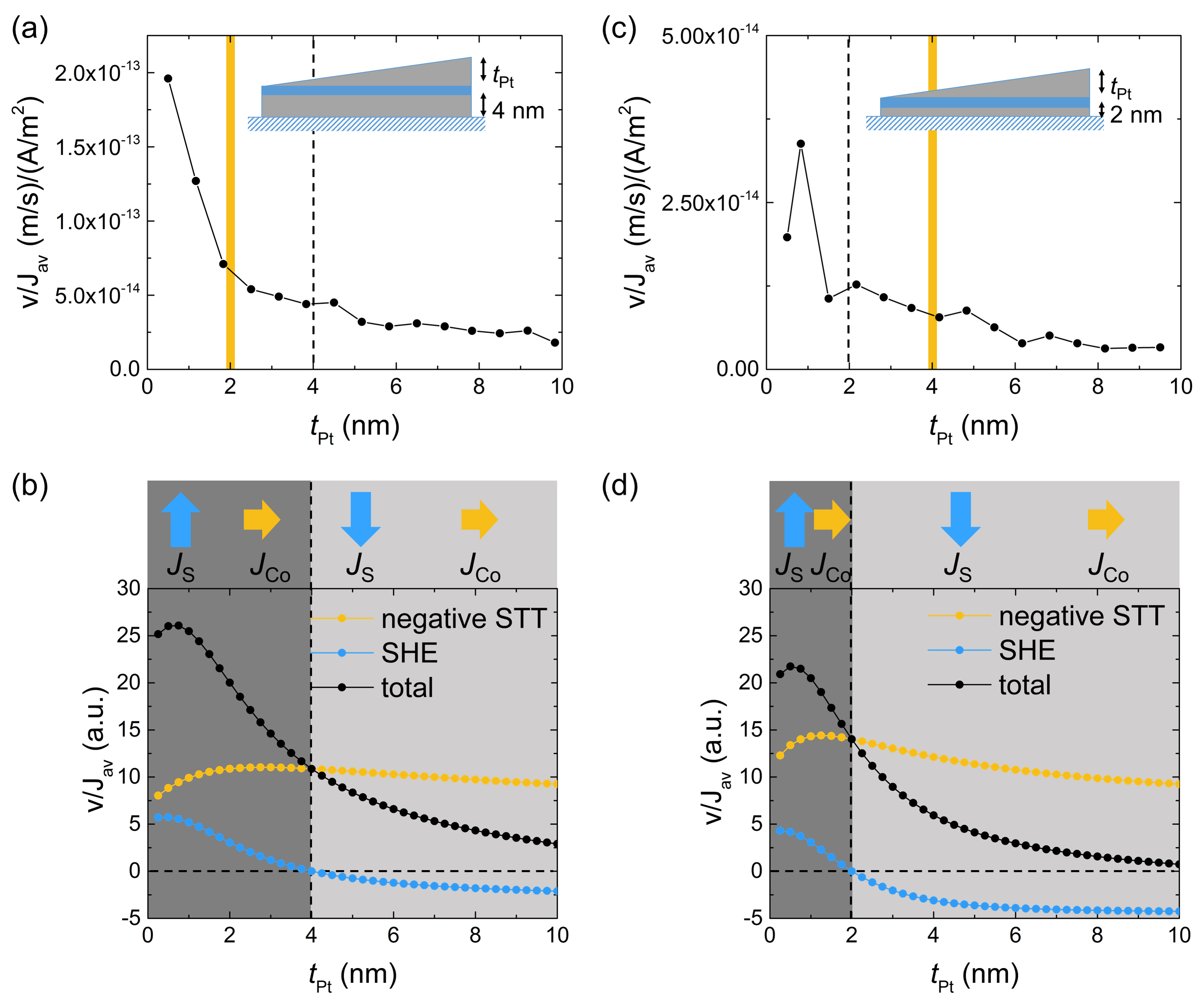}
\caption{Velocity per (average) current density as a function of the top Pt layer thickness for the sample with (a) a 4 nm thick bottom Pt layer (c) a 2 nm thick bottom Pt layer. (b+d) Calculation of the layer specific current densities using the Fuchs-Sondheimer model for the same material stacks which are used experimentally.}
\label{C6:fig4}
\end{figure}


Experimentally we find that $v$ is proportional to $J_\textrm{av}$ in all our measurements. Using the one dimensional model for DW motion, we find that the velocity as a result of the STT (either positive or negative) is proportional to the current density in the magnetic layer, $J_{\textrm{Co}}$, and the velocity as a result of the SHE is proportional to the net injected spin Hall current density, $J_{\textrm{S},net}$, see supplementary material. Here pinning and thermal effects are not considered and current densities between 0 and $2.0 \times 10^9$ A/m$^2$ are investigated (which corresponds to the current densities that are used experimentally). Moreover, in case both effects are present, it is found that the velocity $v$ is equal to the sum of the velocities that would have been caused by the individual contributions, so $v=C_1 J_{\textrm{Co}}+C_2 J_{\textrm{S},net}$, where $C_1$ and $C_2$ are some proportionality constants. Dividing by the average current density in the whole stack, $J_{\textrm{av}}$, transforms this equation to the form: $v/J_{\textrm{av}}=C_1 J_{\textrm{Co}}/J_{\textrm{av}}+C_2 J_{\textrm{S},net}/J_{\textrm{av}}$. The benefit of this form is that $v/J_{\textrm{av}}$ is experimentally accessible (for each Pt thickness, the velocity per average current density, $v/J_{\textrm{av}}$, can be determined by a linear fit through the data points, and the results are plotted in Fig.\ \ref{C6:fig4}(a)), while $J_{\textrm{Co}}/J_{\textrm{av}}$ and $ J_{\textrm{S},net}/J_{\textrm{av}}$ can be estimated using the Fuchs-Sondheimer model. Calculations using parameters from Cormier \textit{et al.} \cite{cormier} and assuming that $\theta_{\textrm{SH}}$ is equal for the the top and bottom Pt layer, are shown in Fig.\ \ref{C6:fig4}(b), where $J_{\textrm{Co}}/J_{\textrm{av}}$ is indicated by the yellow points and $J_{\textrm{S},net}/J_{\textrm{av}}$  by the blue points. For $J_{\textrm{Co}}/J_{\textrm{av}}$ we see only a small variation with Pt thickness, related to scattering at the outer surfaces. For $J_{\textrm{S},net}/J_{\textrm{av}}$ a different behaviour is observed, as expected. There is a zero crossing when the top Pt layer has the same thickness as the bottom layer (indicated by the dashed, gray line). The values for $C_1$ and $C_2$ are not trivial to calculate: one can imagine that they depend on several parameters that are difficult to access. Examples are the spin Hall angle, the configuration of the DW (Bloch, N\'{e}el, or something in between), the spin polarization, the damping parameter and the nonadiabaticity, $\beta$ (in the supplementary material it is shown how these parameters appear in the one dimensional model for DW motion). Moreover, the general view is that the bulk STT becomes smaller for thinner magnetic layers \cite{ono,cormier}. Why this effect would become large and negative for atomically thin layers, as was indicated by the experiments of Je \textit{et al.} is unclear, which makes it impossible to calculate how large this contribution is. We therefore empirically try to add the contributions together for different values of $C_1$ and $C_2$. It turns out that when taking $C_2=1.5 C_1$ (the resulting curve is shown in black), the qualitative behaviour of the experimental data can be reproduced. This includes the fact that the skyrmion bubble velocity is always in the direction of the charge current for the layer thicknesses investigated and that the velocity decreases for larger top Pt layer thicknesses.

Theoretically it is expected that the direction of the spin current injected into the Co layer because of the SHE depends on the thicknesses of the top and bottom Pt layers. However, we have only proven experimentally that this direction is opposite for two thickness combinations: Pt (4 nm) / Co / Pt (2 nm) and Pt (2 nm) / Co / Pt (4 nm) (see again supplementary material). Only one of these specific combinations occurs in our double wedge sample, as it has a constant bottom Pt thickness. For completeness and as an additional test, we therefore also grow a double wedge sample with a bottom Pt layer of 2 nm. The measured results are shown in Fig.\ \ref{C6:fig4}(c). The positions with Pt thickness combinations for which we have checked that the net spin current has the opposite direction are indicated by the vertical yellow lines in Fig.\ \ref{C6:fig4}(a) and (c). In both cases, a positive velocity is measured, meaning that although the spin current is opposite, the skyrmion motion is in the same direction. Please note that this could still be explained by the SHE alone in case the DMI has an opposite sign in the two samples, but additional experiments presented in the supplementary material render this explanation unlikely. The results on the double wedge with a bottom Pt layer of 2 nm turn out to be very similar to the results for the sample with the 4 nm Pt layer: there is no point at which the direction of motion reverses and all observed velocities are positive and the observed velocities per current density are larger for thin top Pt layers than for thick top Pt layers. In Fig.\ \ref{C6:fig4}(d) calculations using the Fuchs-Sondheimer model are shown, using the same material parameters as in Fig.\ \ref{C6:fig4}(b), except for that now the bottom Pt layer thickness is set to 2 nm. Again, there is a qualitative agreement between the calculations and the experimental results. All this suggests that a combination of the SHE and a negative bulk STT is a plausible explanation for the observed skyrmion bubble motion.

\section{Discussion and outlook}

The results in this paper show that additional current-induced effects, besides spin-orbit torques and traditional bulk spin-transfer torques, can play a role in skyrmionic systems. The most likely explanation for our observation is the existence of a negative bulk spin-transfer torque, for which there have previously been indications in a study on DWs in atomically thin Co films \cite{negativeSTT}. There are various sorts of bulk STTs: there is an adiabatic and a non-adiabatic term, as well as higher order contributions. We will now briefly discuss how likely these terms are to cause the negative STT we observe. The higher order contributions become increasingly important for narrow DWs, which might explain why the effect is, up to now, only observed in atomically thin films. However, the underlying physical mechanism for this contribution is momentum transfer, which should never result in motion against the electron flow \cite{Tatara2004}. For the standard adiabatic and non-adiabatic terms it is theoretically possible to drive a domain wall or skyrmion against the electron flow, when the spin polarization in the magnetic layer or the nonadiabaticity becomes negative \cite{Gilmore2011}. However, why these bulk effects would dominate over the SOT induced effects in atomically thin layers remains unclear. Therefore, the origin of this negative STT is a subject on which further investigation is required. Using giant magnetoresistance measurements we have searched for signs of a negative spin polarization, but so far non were found, see supplementary material. 

Irrespective of the precise origin of the observed effect, there are some practical implications. As mentioned before, the DMI is intensively investigated, but remains difficult to measure. Because it is directly related to the chirality of the skyrmions it helps to stabilize, the direction of skyrmion motion under influence of the SHE reveals the sign of the DMI. When skyrmion motion is observed \textemdash especially in the direction against the electron flow for which up to now no other mechanism than the SHE was considered likely \textemdash its direction is often used to deduce the sign of the DMI and compare this with the expectation for the used material stack. Now knowing that there is an alternative effect that can lead to skyrmion motion against the electron flow, this type of analysis becomes unreliable. We would therefore advise to always do additional measurements to determine the chirality of skyrmions if this is of importance, such as direct imaging of the magnetization orientation inside the DWs or checking the presence of the skyrmion Hall effect. 

\bibliography{library}

\section{Acknowledgements}

This work is part of the research programme of the Foundation for Fundamental Research on Matter (FOM), which is part of the Netherlands Organisation for Scientific Research (NWO).

\section{Author contributions}

F.U. conceived the experiment, conducted the experiment and analysed the results. T.W. and T.L. were involved in the sample fabrication and measurements. R.D. provided fruitful discussions about which theoretical phenomena could explain the experimental results. R.L. provided the computer code for the Fuchs-Sondheimer calculations and performed prior experiments that motivated this research. H.S. supervised the study. All authors reviewed the manuscript.

\section{Competing interests}
The authors declare no competing interests.

\section{Additional information}
Supplementary material is available online.

The data on which the findings of this study are based are available from the corresponding author upon reasonable request. Correspondence and requests for materials
should be addressed to F.U.

\section{Methods}
The material stacks used in this work are all grown on thermally oxidized SiO$_2$ substrates. Before the deposition, the substrates are cleaned using acetone and isopropanol in a ultrasonic bath. For further cleaning, they are exposed to an oxygen plasma for 10 minutes. Deposition is done by d.c. sputtering using a argon plasma of approximately 0.01 mbar, in a system with a base pressure of approximately $10^{-7}$ mbar. The thickness variations/wedge geometries are achieved by gradually covering the sample with a mask during the deposition.

The data in Fig. 1(b) was obtained from hysteresis loops that are measured using a polar MOKE setup. All other measurements are performed using a Kerr microscope from Evico magnetics \cite{kerrmicroscope}. A Picoprobe tungsten probe tip with a diameter of 35 micrometer and a Keithley 2400 SourceMeter are used for current injection. The data is analysed using a home-made MATLAB script, of which the principles are explained in the supplementary material.

\end{document}